\documentclass[superscriptaddress,twocolumn,nofootinbib,showpacs,pra,amsmath]{revtex4}
\usepackage{times}
\usepackage{graphicx}
\usepackage{longtable}
\usepackage{color}
\usepackage{amsmath}
\usepackage{longtable}
\usepackage{amssymb}

\newcommand{\beq}{\begin{equation}}
\newcommand{\eeq}{\end{equation}}
\newcommand{\ket} [1] {\vert#1\rangle}
\newcommand{\bra} [1] {\langle#1\vert}

\newcommand{\LC}{\ket{{\rm LC}_6}}
\newcommand{\HE}{\ket{{\rm HE}_6}}
\newcommand{\CZ}{\textsf{CZ}}
\newcommand{\CX}{\textsf{CX}}
\newcommand{\HH}{\textsf{H}}
\newcommand{\LCtilde}{\ket{\widetilde{\rm LC}_6}}
\newcommand{\HEtilde}{\ket{\widetilde{\rm HE}_6}}

\begin{document}

\title{Six-qubit two-photon hyperentangled cluster states: characterization and application to quantum computation}
\author{Giuseppe Vallone}
\homepage{http://quantumoptics.phys.uniroma1.it/}
\affiliation{Museo Storico della Fisica e Centro Studi e Ricerche Enrico Fermi, Via
Panisperna 89/A, Compendio del Viminale, Roma 00184, Italy}
\affiliation{Dipartimento di Fisica, 
Universit\`{a} Sapienza di Roma,
Roma 00185, Italy}
\author{Gaia Donati}
\homepage{http://quantumoptics.phys.uniroma1.it/}
\affiliation{Dipartimento di Fisica, 
Universit\`{a} Sapienza di Roma,
Roma 00185, Italy}
\author{Raino Ceccarelli}
\homepage{http://quantumoptics.phys.uniroma1.it/}
\affiliation{Dipartimento di Fisica, 
Universit\`{a} Sapienza di Roma,
Roma 00185, Italy}
\author{Paolo Mataloni}
\homepage{http://quantumoptics.phys.uniroma1.it/}
\affiliation{Dipartimento di Fisica, 
Universit\`{a} Sapienza di Roma,
Roma 00185, Italy}
\affiliation{Istituto Nazionale di Ottica Applicata (INOA-CNR), L.go E. Fermi 6, 50125 Florence, Italy}

\date{\today}


\begin{abstract}
Six-qubit cluster states built on the simultaneous entanglement of two photons in three 
independent degrees of freedom, i.e. polarization and a double longitudinal momentum, 
have been recently demonstrated. We present here the peculiar entanglement properties of
 the linear cluster state $\LCtilde$ related to the three degrees of freedom. This
 state has been adopted to realize various kinds of Controlled NOT (\textsc{Cnot}) gates, obtaining
 in all the cases high values of the gate fidelity. Our results demonstrate that
 a number of qubits $\leq$10 in cluster states of two photons entangled in multiple degrees
 of freedom is achievable. Furthermore, these states represent a promising approach 
towards scalable quantum computation in a medium term time scale. The future perspectives
 of a {\it hybrid} approach to one-way quantum computing based on multi-degree of freedom and multi-photon
 cluster states are also discussed in the conclusions of this paper. 

\end{abstract}



\maketitle


\section{INTRODUCTION}
Multiqubit graph states \cite{hein04pra} are a basic resource for a number of important quantum information applications. 
These states have been proposed in particular for advanced tests of quantum nonlocality in which the 
violation of local realism increases exponentially with the number of qubits \cite{merm90prl,guhn05prl,cabe07prl,cabe08pra},  and for
 the realization of quantum computation algorithms of increasing complexity in the one-way model \cite{raus01prl,brie01prl}.
 Other application fields deal with quantum communication \cite{clev99prl} and quantum error correction \cite{schl01pra}. 

In recent years, photon cluster states of four, six and up to ten qubits have been realized
 by different approaches and used to deeply investigate the peculiar properties of
 high dimensional entanglement \cite{brie09nap} and to perform basic quantum computation algorithms \cite{chen07prl,vall08prl}. 

\begin{figure}[t]
\begin{center}
\includegraphics[width=9cm]{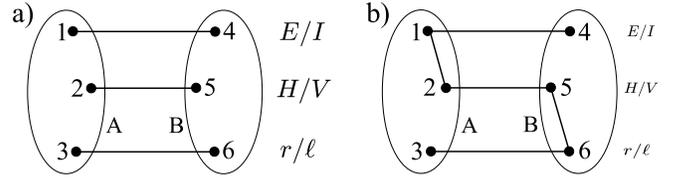}
\caption{(a) Graph associated to the hyperentangled
state $\HE$. Each set represents a photon and every vertex
is associated to a qubit. Qubits $1$ and $4$ are encoded
in the $E/I$ DOF, qubits $2$ and $5$ in polarization and qubits
$3$ and $6$ in the $r/\ell$ DOF. See text for further details. (b)
Graph associated to the two-photon six-qubit linear cluster state
$\LC$. $\LC$ can be obtained from $\HE$ by application of two \CZ\ 
 operations between qubits belonging to different DOFs.}
\label{HE:LC6}
\end{center}
\end{figure}
Two strategies are generally used to create multiqubit cluster states: one 
consists of increasing the number of entangled photons \cite{zhao03prl,walt05nat,walt05prl,kies05prl,prev07nat}, the 
second one is based on the encoding of more qubits in different degrees of
 freedom of the particles \cite{chen07prl,vall08prl,vall07prl,gao08qph}. By the first approach, some 
examples of four and six photon \cite{zhao03prl,walt05nat,walt05prl,kies05prl} cluster states have 
been experimentally demonstrated, up to now, with very
 low rates. The second approach, which is based on two-photon 
hyperentanglement, has been used to create two-photon four-qubit
 cluster states \cite{vall07prl,gao08qph,toku08prl,mair01nat,cine05prl,barb05pra,barr05prl,schu06prl,park07ope,lany09nap,vall09pra}. 
By using 
hyperentanglement, five photons have been recently entangled
 in ten qubits encoded in the polarization and longitudinal 
momentum degrees of freedom (DOFs) \cite{gao08qph}. 

The advantages of the hyperentangled state approach, as 
far as generation/detection rate and fidelity of the states
 are concerned, have been already demonstrated \cite{chen07prl,vall08prl}.
 These properties have been very recently confirmed by
 the realization of the linear 2-photon 6-qubit cluster 
state $\LCtilde$ starting from the triple entanglement of two 
photons in three independent DOFs \cite{cecc09prl}, namely the
 polarization and a double longitudinal momentum. The 
$\LCtilde$ is the only distribution of six qubits between two 
particles whose perfect correlations have the same nonlocality
 as those of the six-qubit Greenberger-Horne-Zeilinger state \cite{cabe08pra},
 but only requires two separated carriers \cite{cabe07prl}.

In this paper we give a detailed characterization of the $\LCtilde$ 
state realized by using the triple hyperentanglement of two
 photons and demonstrate its feasibility for one-way quantum 
computation by the high fidelity realization of different kinds
 of \textsc{Cnot} gates. 

The paper is organized as follows. In Sec. \ref{sec:generation} we describe
 the realization of the six-qubit linear cluster state, derived
 from the application of suitable \textsc{Cphase} gates to a six-qubit 
hyperentangled state. Sec. \ref{sec:char} reports on the  characterization 
of the $\LCtilde$ state by a sequence of quantum tomographic 
reconstructions performed in the three DOFs. Sec. \ref{sec:CNOT} describes
 how the \textsc{Cnot} gate has been efficiently realized with six qubits.
 Finally, the future perspectives of the realization of multiqubit
 cluster states built on an increasing number of photon DOFs
 are discussed in the conclusions of Sec. \ref{sec:concl}.

\section{GENERATION OF THE SIX-QUBIT CLUSTER STATE}
\label{sec:generation}
\begin{figure*}[t]
\begin{center}
\includegraphics[width=16cm]{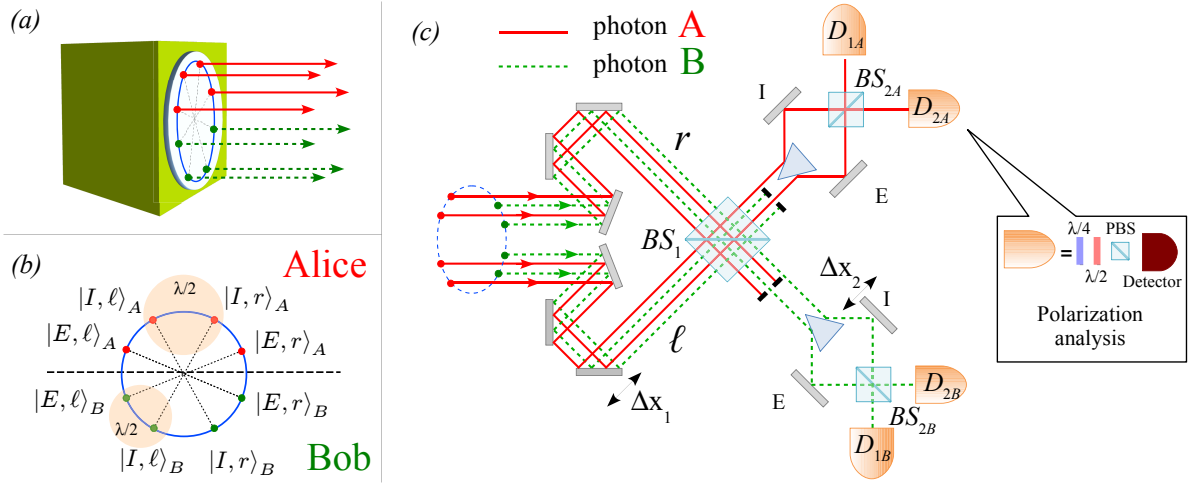}
\caption{Setup of the experiment. (a) Source (green box) of the 8-mode hyperentangled 
state. A detailed description of the source
is given in \cite{barb05pra,cine05prl,cecc09prl}. 
(b) Mode labelling: upper (lower) modes correspond to Alice (Bob) photon. 
For each photon we indicate 
with $\ket r$ ($\ket\ell$) the right (left) modes
and with $\ket I$ ($\ket E$) the internal (external) modes. 
We also show the two half wave plates ($\lambda/2$)
used to transform the hyperentangled state $\HEtilde$ to 
the cluster state $\LCtilde$. The $\lambda/2$ on the $I$ modes of photon $A$
is oriented at $45^\circ$ while the $\lambda/2$ on the $\ell$ modes of photon $B$
is oriented at $0^\circ$. 
(c) Measurement scheme: the momentum measurement setup consists of two chained interferometers,
the first ($BS_1$) measuring the $r/\ell$ qubit, while the second ($BS_{2A}$ and $BS_{2B}$)
measuring the $I/E$ qubit. Polarization analysis is performed by standard waveplates and 
polarizing beam splitters (PBS). Path delays $\Delta x_1$ and $\Delta x_2$, are varied to
obtain the optimal temporal superposition of the modes respectively in the first and second interferometer.
}
\label{fig:setup}
\end{center}
\end{figure*}
Cluster states are peculiar entangled states associated to
$n$-dimensional lattices where each vertex $i$ represents a qubit
and connections between vertices correspond to Ising interactions
between the two-level quantum systems. Two-dimensional lattices
have proved to be a universal resource for Quantum Computation (QC) \cite{raus01prl}; from here on, we
shall then restrict ourselves to the case $n = 2$. The explicit
expression of a cluster state is obtained by preparation of each
qubit in the state $\ket{+}_i = \frac{1}{\sqrt{2}}(\ket{0}_i +
\ket{1}_i)$ and subsequent application of a \textsc{Cphase} gate,
$\CZ_{ij}$, between two adjacent vertices $i$ and $j$. We have
\begin{equation}
\CZ_{ij} = \ket{0}_i \bra{0} \otimes \openone_j + \ket{1}_i \bra{1}
\otimes Z_j,
\end{equation}
where $\openone$ is the identity operator. From now we will use the following
simplified notation for the Pauli operators: $\sigma_{z}^{(i)} \equiv
Z_i$ and analogous relations for $\sigma_{x}^{(i)}$ and
$\sigma_{y}^{(i)}$.

For a lattice $\mathcal L$ with $N$ sites, the corresponding
cluster state can then be written as
\begin{equation}\label{eq:defcluster}
\ket{\Phi_{N}^{\mathcal{L}}} = \Bigl(\prod_{i,j \,
\mathrm{linked}} \CZ_{ij}\Bigr)\ket{+}^N,
\end{equation}
where $\ket{+}^N = \ket{+}_1 \otimes \ket{+}_2 \otimes \ldots
\otimes \ket{+}_N$.

In general, the cluster state associated to a specific graph can
be equivalently defined as the only state satisfying the eigenvalue
equations
\begin{equation}
g_i \ket{\Phi_{N}^{\mathcal{L}}} = \ket{\Phi_{N}^{\mathcal{L}}}
\end{equation}
for every lattice vertex $i$, where the operators
\begin{equation}\label{eq:stabilizer}
g_i = X_i \bigotimes_{j \in \mathcal{N}_i} Z_j
\end{equation}
are known as the stabilizer generators for the cluster state. 
$\mathcal{N}_i$ is the set of vertices connected with the vertex $i$.

The linear cluster state $\LC$ is the state associated to the lattice
shown in Fig. \ref{HE:LC6}(b). We generated a six-qubit two-photon linear cluster state
$\LCtilde$, equivalent to $\LC$ up to single qubit unitary transformations, starting
 from the hyperentangled state $\HEtilde$
and exploiting the three degrees of freedom (DOFs) of polarization and
two different kinds of longitudinal momentum.
To show that the cluster state $\LCtilde$ obtained in the
laboratory is equivalent to $\LC$, we start
describing the source of the hyperentangled state $\HEtilde$, 
the first step for the generation of the linear cluster $\LCtilde$.

The two-photon six-qubit source,
extensively described elsewhere
\cite{barb05pra,cine05prl,cecc09prl}, consists of a continuous wave
(cw), vertically-polarized Ar$^{+}$ laser beam 
($P = 50 mW$, $\lambda_{p} = 364 \, nm$) interacting through spontaneous
parametric down-conversion (SPDC) with a Type I, 0.5 $mm$ thick
$\beta$-Barium-Borate (BBO) crystal. The nonlinear interaction between the
laser beam and the BBO crystal produces degenerate photon pairs at
wavelength $\lambda = 728 \, nm$, entangled in polarization and belonging to the surfaces of an
emission cones. Referring to Fig. \ref{fig:setup}(a),
the insertion of a holed mask allows us to select four pairs of
correlated spatial modes from the conical surface, which is all we
need for the creation of the hyperentangled state $\HEtilde$. The
labels used to identify the selected modes require some
explanations [cfr. Fig. \ref{fig:setup}(b)]: the distinction between left and right modes
provides us with the first longitudinal momentum DOF ($r/\ell$,
also known as the linear momentum $\mathbf{k}$), while
distinguishing between external and internal modes supplies the
second momentum DOF ($E/I$). Moreover, the conical emission of the
BBO crystal can be divided into an ``up'' circular half and a
``down'' one with respect to an ideal horizontal line passing
through the center of the mask. Every mode belonging to the ``up''
half shall be associated to carrier photon $A$; an analogous
correspondence is adopted for the ``down'' half and the second
carrier photon $B$. By doing so we have at our disposal two SPDC
photons, $A$ and $B$, to each of which we associate three
different qubits corresponding to the three DOFs 
(polarization, first, and second momentum) introduced above.

By appropriately setting the phase of each pair of modes, 
the source generates the hyperentangled state $\HEtilde$, explicitly written as
\begin{widetext}
\begin{equation}\label{eq:hyperent}
\begin{aligned}
\HEtilde &= \frac{1}{\sqrt{2}}({\ket{HH}}_{AB} - {\ket{VV}}_{AB}) \otimes
\frac{1}{2}(
{\ket{Er}}_A{\ket{E\ell}}_{B} +
{\ket{E\ell}}_A{\ket{Er}}_{B} +
{\ket{Ir}}_A{\ket{I\ell}}_{B} +
{\ket{I\ell}}_A{\ket{Ir}}_{B} 
) = 
\\
&= \frac{1}{\sqrt{2}}({\ket{EE}}_{AB} + {\ket{II}}_{AB})
\otimes \frac{1}{\sqrt{2}}({\ket{HH}}_{AB} - {\ket{VV}}_{AB}) \otimes
\frac{1}{\sqrt{2}}({\ket{r \ell}}_{AB} + {\ket{\ell r}}_{AB})
\end{aligned}
\end{equation}
\end{widetext}
It comes out that the state $\HEtilde$  
is given by a tensor product of three maximally entangled state, one for each DOF.

By setting the following correspondences between physical and
computational qubits,
\begin{subequations}
\label{eq:whole}
\begin{eqnarray}
\{\ket{E}_A, \ket{I}_A\} &\rightarrow& \{\ket{0}_1, \ket{1}_1\}, \label{notation:1} \\
\{\ket{H}_A, \ket{V}_A\} &\rightarrow& \{\ket{0}_2, \ket{1}_2\}, \\
\{\ket{r}_A, \ket{\ell}_A\} &\rightarrow& \{\ket{0}_3,
\ket{1}_3\},
\\ \{\ket{E}_B, \ket{I}_B\} &\rightarrow& \{\ket{0}_4,
\ket{1}_4\}, \\ \{\ket{H}_B,
\ket{V}_B\} &\rightarrow& \{\ket{0}_5, \ket{1}_5\}, \\
\{\ket{r}_B, \ket{\ell}_B\} &\rightarrow& \{\ket{0}_6,
\ket{1}_6\}, \label{notation:2}
\end{eqnarray}
\end{subequations}
we can express the state \eqref{eq:hyperent} as
\begin{equation}
\HEtilde = \HH_2 X_3 \HH_3 \HH_4 Z_5 \HE,
\end{equation}
where $\HE$ is the state associated to the graph shown in Fig. \ref{HE:LC6}(a) and
$\HH_i$ is the Hadamard operator acting on qubit $i$. 
From the definition of graph states in eq. \eqref{eq:defcluster}, $\LC$ is obtained from the graph state $\HE$ by the application of the
two-qubit gates $\CZ_{12}$ and $\CZ_{56}$. 

We build the state $\LCtilde$ 
by applying the gates $\CX_{12}$ and $\CZ_{56}$ to the hyperentangled state $\HEtilde$. 
The gate $\CX$ is defined as $\CX_{ij} = \ket{0}_i \bra{0} \otimes \openone_j + \ket{1}_i \bra{1}\otimes X_j$.
We are now in
the position to state the relation between the state $\LCtilde$,
and the state $\LC$:
\begin{equation}\label{eq:clusterequiv}
\begin{aligned}
\LCtilde &=  \CX_{12} \, \CZ_{65} \HEtilde\\
& =    \CX_{12} \, \CZ_{65} (\HH_2 X_3 \HH_3 \HH_4 Z_5)\HE\\
& =   (\HH_2 X_3 \HH_3 \HH_4 Z_5) \CZ_{12} \, \CZ_{65}\HE\\
&=  \HH_2 X_3 \HH_3 \HH_4 Z_5 \LC.
\end{aligned}
\end{equation}
The previous relations can be easily demonstrated by using the property $\CX_{ij}\HH_j=\HH_j\CZ_{ij}$.
We thus see that the generated cluster state $\LCtilde$ is
equivalent to the linear six-qubit two-photon cluster state $\LC$
up to the unitary transformation $[\HH_2 X_3 \HH_3 \HH_4 Z_5]$ consisting of single qubit unitaries.
In the generated state, qubits
$1$ and $4$ are encoded in the $E/I$ longitudinal momentum DOF,
qubits $2$ and $5$ in the polarization variable and qubits $3$ and
$6$ in the $r/\ell$ momentum DOF (see Fig. \ref{HE:LC6}).
Specifically, the relation given in \eqref{eq:clusterequiv}
between $\LCtilde$ and $\LC$ implies that $\LCtilde$ is the only
common eigenstate of the generators $\{\widetilde{g}_i\}$ obtained
from $\{g_i\}$ by changing $X_2 \leftrightarrow Z_2$, $X_3
\rightarrow -Z_3$, $Z_3 \rightarrow X_3$, $X_4 \leftrightarrow
Z_4$ and $X_5 \rightarrow -X_5$.

Starting from Eq. \eqref{eq:clusterequiv}, we can write the
following explicit expressions for the generated state $\LCtilde$
by differently factoring the terms referring to the three
considered DOFs:
\begin{widetext}
\begin{subequations}\label{eq:equiv}
\begin{align}\label{eq:equiv1}
\LCtilde &= \frac{1}{2} \bigl[\ket{EE}\ket{\phi^{+}}_{\pi}\ket{r
\ell} + \ket{EE}\ket{\phi^{-}}_{\pi}\ket{\ell r} +
\ket{II}\ket{\psi^{+}}_{\pi}\ket{r \ell} -
\ket{II}\ket{\psi^{-}}_{\pi}\ket{\ell r}\bigr] =
\\\label{eq:equiv2}
&= \frac{1}{2} \bigl[\ket{EE}\ket{HH}\ket{\psi^{+}}_{\mathbf{k}} +
\ket{EE}\ket{VV}\ket{\psi^{-}}_{\mathbf{k}} +
\ket{II}\ket{VH}\ket{\psi^{+}}_{\mathbf{k}} +
\ket{II}\ket{HV}\ket{\psi^{-}}_{\mathbf{k}}\bigr] = \\\label{eq:equiv3}
&= \frac{1}{2} \bigl[\ket{\phi^{+}}_c\ket{++}\ket{r \ell} +
\ket{\phi^{-}}_c\ket{--}\ket{r \ell} +
\ket{\phi^{+}}_c\ket{+-}\ket{\ell r} +
\ket{\phi^{-}}_c\ket{-+}\ket{\ell r}\bigr],
\end{align}
\end{subequations}
\end{widetext}
where we omitted the subscripts $AB$. The states
$\ket{\phi^{\pm}}_{\pi} = \frac{1}{\sqrt{2}} \bigl(\ket{HH}_{AB}
\pm \ket{VV}_{AB}\bigr)$ and $\ket{\psi^{\pm}}_{\pi} =
\frac{1}{\sqrt{2}} \bigl(\ket{HV}_{AB} \pm \ket{VH}_{AB}\bigr)$
are the four polarization Bell states, while the states
$\ket{\psi^{\pm}}_{\mathbf{k}}$ and $\ket{\phi^{\pm}}_c$ are the
standard Bell states encoded in the $r/\ell$ and $E/I$ degrees of
freedom, respectively (the ``$c$'' subscript standing for ``cone'').

The realization of the two-qubit gates responsible for the
transformation of the hyperentangled state $\HEtilde$ into the
cluster state $\LCtilde$ in terms of optical components was made
possible by the insertion of two wave-plates after the holed mask;
since qubits $1$ and $2$ belong to photon $A$, the first $\CX_{12}$
gate was realized by means of a $\lambda/2$ wave-plate oriented at
$45^{\circ}$ and intercepting the two internal $A$ modes (see Fig. \ref{fig:setup}(b)
 and Eq. (\ref{notation:1})). Analogously, the $\CZ_{65}$
gate was obtained thanks to a second $\lambda/2$ wave-plate
oriented at $0^{\circ}$ and intercepting the two left $B$ modes
(see Fig. \ref{fig:setup}(b) and Eq. (\ref{notation:2})). It actually proved
convenient to have two separated $\lambda/2$ wave-plates on the
left $B$ modes, but this was a choice uniquely related to our
specific experimental setup.

\section{CHARACTERIZATION OF THE SIX-QUBIT CLUSTER STATE}
\label{sec:char}
Let us refer to Fig. \ref{fig:setup}(c). The two chained interferometers,
whose core elements are the three symmetric beam splitters
$BS_1$, $BS_{2A}$ and $BS_{2B}$, allow the simultaneous
measurement of the three single-qubit compatible observables
associated to both particles $A$ and $B$. The $r$ modes are made
indistinguishable (in space as well as in time) from the $\ell$
ones on $BS_1$, while $E$ and $I$ modes are matched on $BS_{2A}$
or $BS_{2B}$ depending on which photon they refer to. By means of
a trombone mirror assembly in each of the two interferometers, it
is possible to act on the optical path delays, $\Delta x_1$ and $\Delta x_2$, 
and find the optimal
temporal superposition conditions for both of the interference
phenomena. Let us now refer to the $BS_1$: we set
$\{\ket{\ell}_j,\ket{r}_j\}$ and
$\{\ket{\ell^{\prime}}_j,\ket{r^{\prime}}_j\}$, for $j = A, B$, as
its input and output states. The insertion of a thin glass
plate intercepting two right $A$ modes (one internal and one external) 
transforms the input states in the following way:
$\ket{\phi_A}_{\mathbf{k}} =
\frac{1}{\sqrt2}(\ket{\ell}_A + e^{-i\phi_A}\ket{r}_A) \rightarrow
\ket{\ell^{\prime}}_A$ and $\ket{\phi^{\bot}_A}_{\mathbf{k}} =
\frac{1}{\sqrt2}(\ket{\ell}_A - e^{-i\phi_A}\ket{r}_A) \rightarrow
\ket{r^{\prime}}_A$, for external and internal modes. By detecting the photons on the 
$\ket{\ell'}$ or the $\ket{r'}$ output we are projecting the input state respectively into 
$\ket{\phi_A}$ or $\ket{\phi^\perp_A}$. An analogous glass plate intercepts the left $B$ 
modes\footnote{In this case the projection is performed into the states
$\ket{\phi_B}_{\mathbf{k}} =
\frac{1}{\sqrt2}( e^{-i\phi_B}\ket{\ell}_B +\ket{r}_B)$ and $\ket{\phi^{\bot}_B}_{\mathbf{k}} =
\frac{1}{\sqrt2}( e^{-i\phi_B}\ket{\ell}_B -\ket{r}_B)$}.

Two more
such phase shifters, intercepting the external A and B modes, are
inserted in the second interferometer before $BS_{2A}$ and
$BS_{2B}$. Four single-photon detectors $D_{1A}$,
$D_{2A}$, $D_{1B}$ and $D_{2B}$ receive the
radiation belonging to the ``up'' and ``down'' output modes (see
Fig. \ref{fig:setup}(b)), which we can label as
$\{\ket{E^{\prime}}_j,\ket{I^{\prime}}_j\}$ for $j = A, B$. In the
presence of the glass plates cited above, the following
input-output transformations concerning $BS_{2A}$ and $BS_{2B}$
hold: $\ket{\delta}_c = \frac{1}{\sqrt2}(e^{-i\delta}\ket{E}_j +
\ket{I}_j) \rightarrow \ket{E^{\prime}}_j$ and
$\ket{\delta^{\bot}}_c = \frac{1}{\sqrt2}(e^{-i\delta}\ket{E}_j -
\ket{I}_j) \rightarrow \ket{I^{\prime}}_j$. Finally, a
polarization analyzer constituted of a $\lambda/2$ wave-plate, a
$\lambda/4$ wave-plate and a polarizing beam splitter (PBS) is
added in front of each detector. In these conditions, we recorded
nearly $500$ coincidences per second.
\begin{table}[b]
\caption{\label{table:tomofidelities}Fidelities of Bell states for each DOF. We selected a separable state for
two DOFs (first column) and performed a tomographic reconstruction of the density matrix of the
remaining DOF (second column). The expected output states and the relative fidelities are shown in
the last two columns.}
\begin{ruledtabular}
\begin{tabular}[c]{|c|ccc|}
separable basis & output DOF & output state & Fidelity \\
\hline
$\ket{EE}_c\ket{r\ell}_{\bf k}$&$\pi$ & $\ket{\phi^{+}}_{\pi}$ & $0.821 \pm 0.014$ \\
$\ket{EE}_c\ket{\ell r}_{\bf k}$&& $\ket{\phi^{-}}_{\pi}$ & $0.917 \pm 0.017$ \\
$\ket{II}_c\ket{r \ell}_{\bf k}$&& $\ket{\psi^{+}}_{\pi}$ & $0.905 \pm 0.013$ \\
$\ket{II}_c\ket{\ell r}_{\bf k}$&& $\ket{\psi^{-}}_{\pi}$ & $0.828 \pm 0.025$ \\
\hline
$\ket{EE}_c\ket{HH}_\pi$&$r/\ell$ & $\ket{\psi^{+}}_{\bf k}$ & $0.897 \pm 0.008$ \\
$\ket{EE}_c\ket{VV}_\pi$&& $\ket{\psi^{-}}_{\bf k}$ & $0.933 \pm 0.016$ \\
$\ket{II}_c\ket{VH}_\pi$&& $\ket{\psi^{+}}_{\bf k}$ & $0.899 \pm 0.009$ \\
$\ket{II}_c\ket{HV}_\pi$&& $\ket{\psi^{-}}_{\bf k}$ & $0.858 \pm 0.017$ \\
\hline
$\ket{++}_\pi\ket{r\ell}_{\bf k}$&$E/I$ & $\ket{\phi^{+}}_c$ & $0.797 \pm 0.015$ \\
\end{tabular}
\end{ruledtabular}
\end{table}
\begin{figure}
\centering
\includegraphics[width=9cm]{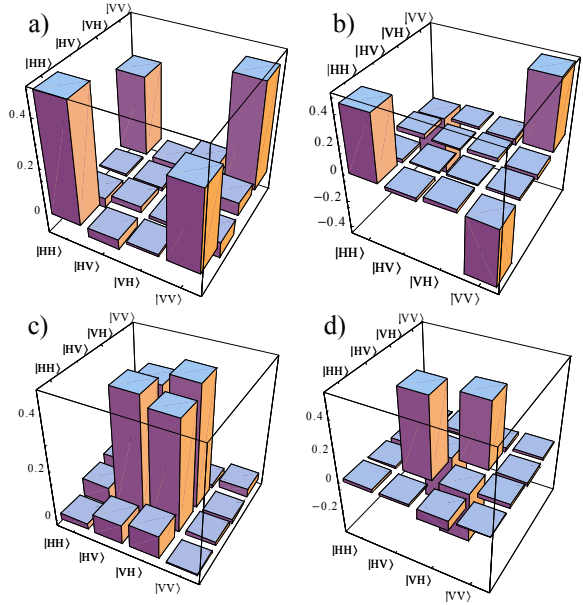}
\caption{Tomographic reconstruction of the four
polarization states in Eq. \eqref{eq:equiv} (real
parts). The imaginary components are negligible. The corresponding
theoretical Bell states are: (a) $\ket{\Phi^{+}}_{\pi}$, (b)
$\ket{\Phi^{-}}_{\pi}$, c) $\ket{\Psi^{+}}_{\pi}$, d)
$\ket{\Psi^{-}}_{\pi}$.} \label{fig:tomo:pi}
\end{figure}
\begin{figure}
\centering
\includegraphics[width=9cm]{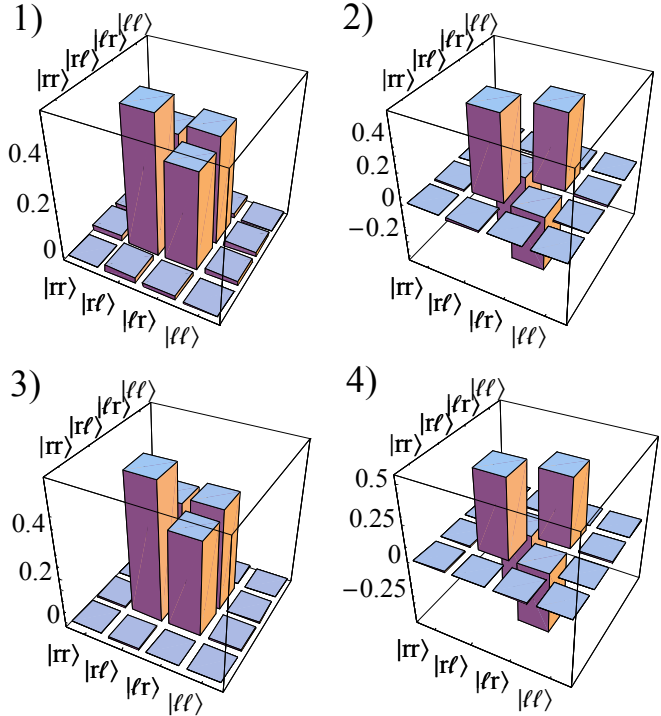}
\caption{Tomographic reconstruction of the four
states encoded in the $r/\ell$ DOF (real parts). The imaginary
components are negligible. The corresponding theoretical Bell
states are: 1) and 3) $\ket{\Psi^{+}}_{\bf k}$, 2) and 4)
$\ket{\Psi^{-}}_{\bf k}$.} \label{fig:tomo:k}
\end{figure}
\begin{figure}
\centering
\includegraphics[width=7cm]{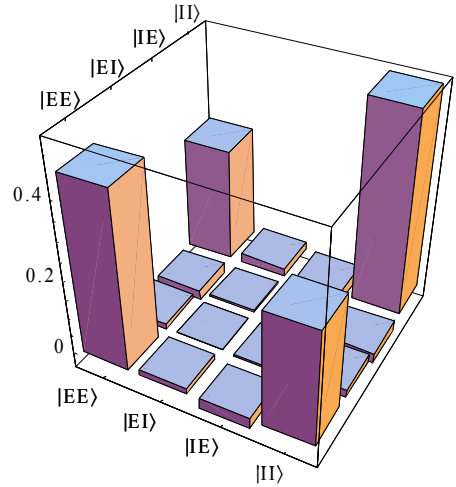}
\caption{Tomographic reconstruction of the state
encoded in the $E/I$ DOF (real part) corresponding to the
$r_A-\ell_B$ spatial mode pair and the polarization state ${\ket{+}}_A{\ket{+}}_B$. The imaginary components is
negligible. The corresponding theoretical state is
$\ket{\phi^{+}}_c$.} \label{fig:tomo:EI}
\end{figure}

The characterization of the generated state $\LCtilde$ relies on a
tomographic reconstruction technique followed by a ``maximum
likelihood'' method \cite{jame01pra}. Particularly, we aim at
recovering Eq. \eqref{eq:equiv}, which shows three alternative and
perfectly equivalent ways of writing the cluster state $\LCtilde$.
Indeed, Eq. \eqref{eq:equiv} is important to prove since it
highlights the inner structure of the generated state.
As we see, each of the expressions \eqref{eq:equiv} is obtained by
writing the states of four qubits corresponding to two DOFs in a separable basis, 
and expressing the remaining couple of qubits
in the appropriate entangled Bell basis; for example, the first relation
shows the four polarization Bell states. Equation \eqref{eq:equiv1} shows that the state $\LCtilde$
is obtained by a coherent superposition between four terms, each of them referring to a specific pair
of correlated modes. We first demonstrated that the four polarization states
 corresponding to the different pairs of modes
are given by the Bell states. The coherence between them can be shown by using equations 
\eqref{eq:equiv2} and \eqref{eq:equiv3}.
It is easy to show that the first two terms in \eqref{eq:equiv2} arise from the superposition
 between the first two terms in \eqref{eq:equiv1},
and the same applies for the last two terms.
By selecting the appropriate separable basis in two DOFs we performed
the tomographic reconstructions to recover the Bell states encoded in
the remaining degree of freedom.
As a consequence, these measurements
prove not only the presence of the various terms appearing in Eq.
\eqref{eq:equiv}, but also implicitly tell us about the coherences
between the states involved.
\begin{table*}[t]
\caption{\label{table:stabilizers}Experimental results: measurement of the
64 stabilizers $\widetilde s_i$ of $\ket{\widetilde{\text{LC}}_6}$, i.e., all the products
of the generators $\widetilde g_i$. Last three columns indicate in which Bell inequality test
each experimental value was used.}
\begin{tabular}[c]{|ccccc|}
\hline \hline
Stabilizer & Experimental value & $\mathcal{B}_{\mathrm{exp}}$ & $\beta$ & $\beta^{\prime}$ \\
\hline
$1$ & $1.0000 \pm 0.0000$ &  &  &  \\
$\tilde g_1$ & $0.5928 \pm 0.0075$ &  & \checkmark &  \\
$\tilde g_2$ & $0.8788 \pm 0.0053$ &  &  &  \\
$\tilde g_3$ & $0.9984 \pm 0.0005$ &  &  &  \\
$\tilde g_4$ & $0.9973 \pm 0.0008$ &  &  &  \\
$\tilde g_5$ & $0.7905 \pm 0.0057$ &  &  &  \\
$\tilde g_6$ & $0.8310 \pm 0.0062$ &  &  & \checkmark \\
$\tilde g_1 \tilde g_2$ & $0.5657 \pm 0.0059$ &  & \checkmark &  \\
$\tilde g_1 \tilde g_3$ & $0.5930 \pm 0.0075$ &  &  &  \\
$\tilde g_1 \tilde g_4$ & $0.5602 \pm 0.0076$ &  & \checkmark &  \\
$\tilde g_1 \tilde g_5$ & $0.5872 \pm 0.0076$ &  &  &  \\
$\tilde g_1 \tilde g_6$ & $0.4653 \pm 0.0095$ & \checkmark &  &  \\
$\tilde g_2 \tilde g_3$ & $0.8586 \pm 0.0062$ &  &  &  \\
$\tilde g_2 \tilde g_4$ & $0.8775 \pm 0.0053$ &  &  &  \\
$\tilde g_2 \tilde g_5$ & $0.7042 \pm 00066$ &  &  &  \\
$\tilde g_2 \tilde g_6$ & $0.8288 \pm 0.0062$ &  &  &  \\
$\tilde g_3 \tilde g_4$ & $0.9970 \pm 0.0009$ &  &  &  \\
$\tilde g_3 \tilde g_5$ & $0.7896 \pm 0.0057$ &  &  &  \\
$\tilde g_3 \tilde g_6$ & $0.7484 \pm 0.0056$ &  &  & \checkmark \\
$\tilde g_4 \tilde g_5$ & $0.7339 \pm 0.0084$ &  &  &  \\
$\tilde g_4 \tilde g_6$ & $0.8312 \pm 0.0062$ &  &  &  \\
$\tilde g_5 \tilde g_6$ & $0.6392 \pm 0.0060$ &  &  & \checkmark \\
$\tilde g_1 \tilde g_2 \tilde g_3$ & $0.4504 \pm 0.0092$ &  &  &  \\
$\tilde g_1 \tilde g_2 \tilde g_4$ & $0.6063 \pm 0.0074$ &  & \checkmark &  \\
$\tilde g_1 \tilde g_2 \tilde g_5$ & $0.5378 \pm 0.0086$ &  &  &  \\
$\tilde g_1 \tilde g_2 \tilde g_6$ & $0.4169 \pm 0.0065$ & \checkmark &  &  \\
$\tilde g_1 \tilde g_3 \tilde g_4$ & $0.5603 \pm 0.0076$ &  &  &  \\
$\tilde g_1 \tilde g_3 \tilde g_5$ & $0.5874 \pm 0.0075$ &  &  &  \\
$\tilde g_1 \tilde g_3 \tilde g_6$ & $0.4651 \pm 0.0063$ & \checkmark &  &  \\
$\tilde g_1 \tilde g_4 \tilde g_5$ & $0.5882 \pm 0.0074$ &  &  &  \\
$\tilde g_1 \tilde g_4 \tilde g_6$ & $0.4148 \pm 0.0075$ & \checkmark &  &  \\
$\tilde g_1 \tilde g_5 \tilde g_6$ & $0.4450 \pm 0.0061$ & \checkmark &  &  \\
 \hline \hline
\end{tabular}
 \qquad
\begin{tabular}[c]{|ccccc|}
\hline \hline
Stabilizer & Experimental value & $\mathcal{B}_{\mathrm{exp}}$ & $\beta$ & $\beta^{\prime}$ \\
\hline
$\tilde g_2 \tilde g_3 \tilde g_4$ & $0.8592 \pm 0.0062$ &  &  &  \\
$\tilde g_2 \tilde g_3 \tilde g_5$ & $0.7036 \pm 0.0066$ &  &  &  \\
$\tilde g_2 \tilde g_3 \tilde g_6$ & $0.7468 \pm 0.0056$ &  &  &  \\
$\tilde g_2 \tilde g_4 \tilde g_5$ & $0.7038 \pm 0.0066$ &  &  &  \\
$\tilde g_2 \tilde g_4 \tilde g_6$ & $0.8285 \pm 0.0062$ &  &  &  \\
$\tilde g_2 \tilde g_5 \tilde g_6$ & $0.6861 \pm 0.0058$ &  &  &  \\
$\tilde g_3 \tilde g_4 \tilde g_5$ & $0.7357 \pm 0.0083$ &  &  &  \\
$\tilde g_3 \tilde g_4 \tilde g_6$ & $0.7484 \pm 0.0056$ &  &  &  \\
$\tilde g_3 \tilde g_5 \tilde g_6$ & $0.6625 \pm 0.0051$ &  &  & \checkmark \\
$\tilde g_4 \tilde g_5 \tilde g_6$ & $0.6394 \pm 0.0060$ &  &  &  \\
$\tilde g_1 \tilde g_2 \tilde g_3 \tilde g_4$ & $0.6067 \pm 0.0074$ &  &  &  \\
$\tilde g_1 \tilde g_2 \tilde g_3 \tilde g_5$ & $0.5391 \pm 0.0086$ &  &  &  \\
$\tilde g_1 \tilde g_2 \tilde g_3 \tilde g_6$ & $0.4334 \pm 0.0063$ & \checkmark &  &  \\
$\tilde g_1 \tilde g_2 \tilde g_4 \tilde g_5$ & $0.4247 \pm 0.0093$ &  &  &  \\
$\tilde g_1 \tilde g_2 \tilde g_4 \tilde g_6$ & $0.3960 \pm 0.0077$ & \checkmark &  &  \\
$\tilde g_1 \tilde g_2 \tilde g_5 \tilde g_6$ & $0.4435 \pm 0.0076$ & \checkmark &  &  \\
$\tilde g_1 \tilde g_3 \tilde g_4 \tilde g_5$ & $0.5897 \pm 0.0074$ &  &  &  \\
$\tilde g_1 \tilde g_3 \tilde g_4 \tilde g_6$ & $0.4349 \pm 0.0080$ & \checkmark &  &  \\
$\tilde g_1 \tilde g_3 \tilde g_5 \tilde g_6$ & $0.4465 \pm 0.0061$ & \checkmark &  &  \\
$\tilde g_1 \tilde g_4 \tilde g_5 \tilde g_6$ & $0.4465 \pm 0.0061$ & \checkmark &  &  \\
$\tilde g_2 \tilde g_3 \tilde g_4 \tilde g_5$ & $0.7037 \pm 0.0066$ &  &  &  \\
$\tilde g_2 \tilde g_3 \tilde g_4 \tilde g_6$ & $0.7465 \pm 0.0056$ &  &  &  \\
$\tilde g_2 \tilde g_3 \tilde g_5 \tilde g_6$ & $0.6113 \pm 0.0063$ &  &  &  \\
$\tilde g_2 \tilde g_4 \tilde g_5 \tilde g_6$ & $0.6860 \pm 0.0058$ &  &  &  \\
$\tilde g_3 \tilde g_4 \tilde g_5 \tilde g_6$ & $0.6624 \pm 0.0051$ &  &  &  \\
$\tilde g_1 \tilde g_2 \tilde g_3 \tilde g_4 \tilde g_5$ & $0.4235 \pm 0.0093$ &  &  &  \\
$\tilde g_1 \tilde g_2 \tilde g_3 \tilde g_4 \tilde g_6$ & $0.3735 \pm 0.0078$ & \checkmark &  &  \\
$\tilde g_1 \tilde g_2 \tilde g_3 \tilde g_5 \tilde g_6$ & $0.4071 \pm 0.0077$ & \checkmark &  &  \\
$\tilde g_1 \tilde g_2 \tilde g_4 \tilde g_5 \tilde g_6$ & $0.5059 \pm 0.0052$ & \checkmark &  &  \\
$\tilde g_1 \tilde g_3 \tilde g_4 \tilde g_5 \tilde g_6$ & $0.4884 \pm 0.0057$ & \checkmark &  &  \\
$\tilde g_2 \tilde g_3 \tilde g_4 \tilde g_5 \tilde g_6$ & $0.6112 \pm 0.0063$ &  &  &  \\
$\tilde g_1 \tilde g_2 \tilde g_3 \tilde g_4 \tilde g_5 \tilde g_6$ & $0.4046 \pm 0.0060$ & \checkmark &  &  \\
\hline \hline
\end{tabular}
 \end{table*}

The reconstruction concerning the
polarization variable exactly followed the strategy presented in
\cite{jame01pra}, while the complete sets of tomographic analysis
states associated to the two longitudinal momentum DOFs were
established combining the known complete set of polarization
states (as given in \cite{jame01pra}) with the stated
correspondence between physical and computational qubits (see
equations \eqref{eq:whole}).

The experimental density matrix reconstructions are shown in Fig.
\ref{fig:tomo:pi} for the polarization variable, in Fig.
\ref{fig:tomo:k} for the linear momentum $\mathbf{k}$ and in Fig.
\ref{fig:tomo:EI} for the $E/I$ DOF. The fidelities associated to
the considered tomographic analysis are listed in Table
\ref{table:tomofidelities}. As we see, most of these values exceed
$80\%$ and some get above $90\%$; the lowest experimental fidelity
corresponds to the tomographic reconstruction associated to the
$E/I$ DOF. 

We attribute this to the difficulty to achieve perfect mode matching 
in the second interferometer due to mode divergences.
Nevertheless, the obtained results represent a first evidence of
the correct generation of the cluster state $\LCtilde$. We also measured the state fidelity 
to give further informations on the state preparation.

As said, the reported tomographic reconstructions allow us to test
the validity of Eq. \eqref{eq:equiv}; this approach is naturally
connected to the first definition of cluster states recalled in
this paper (see Eq. \eqref{eq:defcluster}).

We can then refer to Eq.
\eqref{eq:stabilizer} instead, which gives the characterization of
cluster states in terms of their stabilizer generators, and adopt
a complementary point of view (with respect to the one condensed
in Eq. (\ref{eq:equiv})) leading to a more complete
 characterization of the cluster state $\LCtilde$. Actually, its
 stabilizer generators $\{\tilde g_i\}_{i=1}^6$ generate the
 so-called stabilizer group
 \begin{equation}
 S(\LCtilde) = \{S_j, j = 1,\ldots,2^6 = 64\} \,, \quad 
S_j=\prod_{i \in I_{j}(G)} \tilde g_i \,,
\end{equation}
where $I_{j}(G)$ is a subset of $\{1,\cdots,6\}$. The $64$
elements $\{S_j\}$ are known as the stabilizing operators of
$\LCtilde$, and satisfy the relation $S_j \LCtilde = \LCtilde
\quad \forall \, j = 1,\ldots,64$.

It can be shown that
\begin{equation}
\LCtilde \bra{\widetilde{\text{LC}}_6} = \frac{1}{64} \sum_{j=1}^{64} S_j
\,.
\end{equation}
The fidelity of the experimental cluster state, whose density
matrix is $\rho_{\mathrm{exp}}$, can then be calculated as
\begin{equation}
F_{\LCtilde} =
Tr[\rho_{\mathrm{exp}}\LCtilde\bra{\widetilde{\text{LC}}_6}] =
\frac{1}{64} \sum_{j=1}^{64} Tr[\rho_{\mathrm{exp}}S_j] \,,
\end{equation}
i.e., by measuring the expectation values of the stabilizing
operators of the generated cluster state. We obtained
$F_{\LCtilde} = 0.6350 \pm 0.0008$. The experimental expectation
values $\{\langle S_j \rangle\}_{j=1}^{64}$ are shown in Table
\ref{table:stabilizers}.

We tested the genuine six-qubit entanglement of the created
cluster state by evaluation of an appropriate entanglement
witness, defined as \cite{toth05prl}
\begin{equation}
\mathcal{W}_F = \openone - 2\LCtilde\bra{\widetilde{\text{LC}}_6} =
\openone - \frac{1}{32} \sum_{j=1}^{64} S_j.
\end{equation}
There is entanglement whenever
\begin{equation}
\langle\mathcal{W}_F\rangle = \openone - 2F < 0 \,.
\end{equation}
We found $\langle\mathcal{W}_F\rangle = -0.270 \pm 0.002$, which
being negative by $135$ standard deviations proves the existence
of a genuine six-qubit entanglement.

 The data present in Table
\ref{table:stabilizers} were also used for a nonlocality test of quantum mechanics \cite{cecc09prl}.
Any local theory in which every single-qubit
Pauli observable can be interpreted as an element of reality as
intended by EPR satisfies the following inequality:
\begin{equation}\label{eq:Bell_ineq}
\mathcal{B} \leq 4 \equiv B_{\mathrm{LHVT}} \,,
\end{equation}
where $\mathcal{B}$ is defined as
\begin{equation}
\mathcal{B} =|g_1(\openone + g_2)(\openone + g_3)(\openone +
g_4)(\openone + g_5)g_6| \,.
\end{equation}
We tested the Bell inequality (\ref{eq:Bell_ineq}) and obtained
$\mathcal{B}_{\mathrm{exp}} = 7.018 \pm 0.028$ (see checked rows
in the third column of Table \ref{table:stabilizers}); this result
implies a degree of nonlocality $\mathcal{D} =
\frac{\mathcal{B}_{\mathrm{exp}}}{B_{\mathrm{LHVT}}}$ equal to
$1.7545 \pm 0.0070$. 
We also tested the persistency of entanglement of $\LCtilde$ 
against the loss of two qubits. This property can be investigated
by considering two alternative Bell inequalities with respect to
(\ref{eq:Bell_ineq}): in the first one qubits $3$ and $6$ are
ignored, while in the second inequality we trace out qubits $1$ and $4$:
\begin{subequations}
\label{simple:Bell}
\begin{align}
\label{simple:Bell:1}
\beta &= |g_1(\openone + g_2)(\openone + g_4)| \stackrel{LHV}{\leq} 2 \,,  \\
\beta^{\prime} &= |(\openone + g_3)(\openone + g_5)g_6|\stackrel{LHV}{\leq}2 \label{simple:Bell:2}
\end{align}
\end{subequations}
By using the measurements given in Table \ref{table:stabilizers} we 
found
\begin{subequations}
\begin{eqnarray}
\beta_{\mathrm{exp}} &=& 2.325 \pm 0.014 \,, \\
\beta^{\prime}_{\mathrm{exp}} &=& 2.881 \pm 0.012 \,,
\end{eqnarray}
\end{subequations}
showing violations of the Bell inequalities
(\ref{simple:Bell:1}) and (\ref{simple:Bell:2}).
See \cite{cecc09prl} for more details concerning Bell inequalities with the 2-photon 6-qubit
cluster state.

\section{EXPERIMENTAL REALIZATION OF THE \textsc{Cnot} GATE}
\label{sec:CNOT}
\begin{figure}[t]
\begin{center}
\includegraphics[width=5cm]{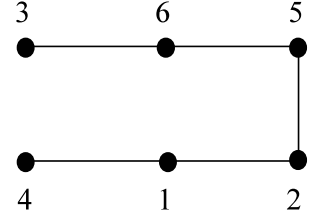}
\caption{Graph associated to the six-qubit horseshoe
cluster state, equivalent to the generated linear cluster state.}
\label{fig:horseshoe}
\end{center}
\end{figure}
Let us now turn to the one-way model of QC \cite{brie09nap}.
Given a cluster state,  it can be useful to think of the
distinct horizontal qubits as ``the original [logical] qubit at different
times'' \cite{niel04prl}, with the temporal axis oriented from
left to right (a choice made possible by appropriately designing
the lattice); single-qubit gates are represented by pairs of
horizontally adjacent qubits, while vertical connections play the
role of \textsc{Cphase} gates. Each computation process is then obtained as
a sequence of single-qubit projective measurements performed on
the so-called physical qubits, simultaneously determining the
propagation of information through the cluster and the loss of
entanglement in the original state \cite{raus01prl,niel04prl}.

This last feature is responsible for the irreversibility of the
process and explains why we speak of one-way computation. The
difference existing between physical and encoded qubits deserves a
deeper understanding. Physical qubits in the initial
cluster state represent an entanglement resource; encoded (or
logical) qubits constitute the quantum information being processed
\cite{raus03pra}. 
\begin{table}[t]
\begin{ruledtabular}
\begin{tabular}[c]{|clcc|}
Pattern& Qubit [DOF] & Measurement CB & Measurement LB \\
\hline
$I$   & $3\,[r/\ell]$ & $\{\ket{0}, \ket{1}\}$       & $\{\ket{+}, \ket{-}\}$ \\
      & $4\,[E/I]$    & $\{\ket{0}, \ket{1}\}$       & $\{\ket{+}, \ket{-}\}$ \\
      & $6\,[r/\ell]$ & $B(\alpha)$\footnotemark[1]  & $B(\alpha)$\footnotemark[1] \\
      & $1\,[E/I]$    & $B(\beta)$\footnotemark[1]   & $B(\beta)$\footnotemark[1] \\
\hline
$II$  & $3\,[r/\ell]$ & $\{\ket{0}, \ket{1}\}$       & $\{\ket{+}, \ket{-}\}$ \\
      & $4\,[E/I]$    & $B(0)$                       & $\{\ket{0}, \ket{1}\}$ \\
      & $6\,[r/\ell]$ & $B(\alpha)$                  & $B(\alpha)$ \\
      & $1\,[E/I]$    & $B(0)$                       & $\{\ket{+}, \ket{-}\}$ \\
\hline
$III$ & $3\,[r/\ell]$ & $B(0)$                       & $\{\ket{1}, \ket{0}\}$ \\
      & $4\,[E/I]$    & $\{\ket{0}, \ket{1}\}$       & $\{\ket{+}, \ket{-}\}$ \\
      & $6\,[r/\ell]$ & $B(0)$                       & $\{\ket{+}, \ket{-}\}$ \\
      & $1\,[E/I]$    & $B(\beta)$                   & $B(\beta)$ \\
\hline
$IV$  & $3\,[r/\ell]$ & $B(0)$       & $\{\ket{1}, \ket{0}\}$ \\
      & $4\,[E/I]$    & $B(0)$       & $\{\ket{0}, \ket{1}\}$ \\
      & $6\,[r/\ell]$ & $B(0)$       & $\{\ket{+}, \ket{-}\}$ \\
      & $1\,[E/I]$    & $B(0)$       & $\{\ket{+}, \ket{-}\}$ \\
\end{tabular}
\end{ruledtabular}
\caption{\label{CNOTcases}Measurement bases for the different considered patterns.
For each pattern we indicate the measured qubit (and the DOF in which the qubit is encoded) and the
corresponding measurement in the cluster (CB) and laboratory bases (LB).}
\label{table:patterns}
\footnotetext[1]{See Eq. (\ref{alphabasis}).}
\end{table}
\begin{figure*}
\centering\includegraphics[width=18cm]{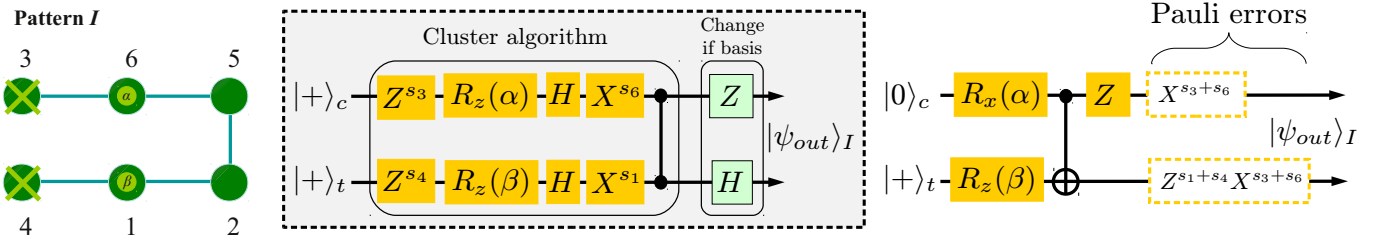}
\caption{Measurement pattern I: we indicate by a cross a measurement 
in the bases $\{\ket0,\ket1\}$ and by $\alpha$ and $\beta$
a measurement in the basis $B(\alpha)$ and $B(\beta)$. The output state
is encoded in qubits 5 and 2. The circuit associated to the
considered measurement is shown. We first indicate the circuit obtained
by directly following the one-way rules and then the equivalent 
circuit composed by single-qubit gates and a two-qubit \textsc{Cnot} gate. The gates
indicated by ''Change of basis'' are due to the change between the computational and
laboratory basis. } 
\label{fig:circuits_caseI}
\end{figure*}
\begin{figure}[t]
\centering\includegraphics[width=9cm]{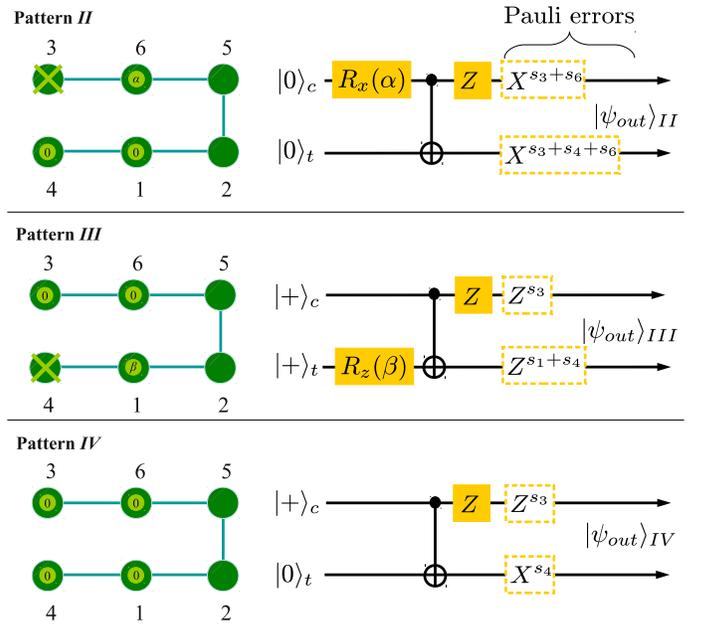}
\caption{Measurement patterns $II$, $III$ and $IV$ and the corresponding
circuit representations. Each circuit is composed by single-qubit 
gates and a two-qubit \textsc{Cnot} gate.} 
\label{fig:circuits}
\end{figure}
Let $N$ be the number of physical qubits and $M$
the number of encoded qubits, with $M < N$. $M$ input cluster
qubits, all prepared in the state $\ket{+}$, are usually
positioned on the left of the two-dimensional graph. The
single-qubit measurements involve $N - M$ qubits. Consequently,
the output of the computation can be read on the $M$ unmeasured
qubits up to local Pauli errors, as will be specified later on in
this paper. More precisely, the measurements driving the
computation are performed in the following basis:
\begin{equation}\label{alphabasis}
B_{i}(\alpha) = \{\ket{\alpha_+}_i, \ket{\alpha_-}_i\},
\end{equation}
with $\ket{\alpha_{\pm}}_i =
\frac{1}{\sqrt{2}}(e^{i\alpha/2}\ket{0}_i +
e^{-i\alpha/2}\ket{1}_i)$. If we take $s_i$ as signalling the
presence of a Pauli error, we usually associate $s_i = 0$ to the
measurement outcome $\ket{\alpha_+}$ (error-free case) and $s_i =
1$ to $\ket{\alpha_-}$. The choice of $\alpha$ (and the consequent
possible errors occurring in the computation) depends on the
algorithm to be implemented. Measuring a qubit in the
computational basis $\{\ket{0}_i, \ket{1}_i\}$ has a completely
different effect on the cluster, in that it removes the measured
qubit and leads to the cluster state
\begin{equation}
\prod_{k \in \mathcal{N}_i} Z^{s_i}_{k}
\ket{\Phi_{N-1}^{\mathcal{L}\backslash\{i\}}},
\end{equation}
where $\mathcal{N}_i$ is the set of vertices connected to site
$i$.

The generated six-qubit cluster allows the implementation of
non-trivial two-qubit operations such as the \textsc{Cnot} gate. For this
purpose, it is convenient to think of a horseshoe ($180^{\circ}$
rotated) six-qubit cluster instead of the one depicted in
Fig.\ref{HE:LC6}(b); the two are physically equivalent, but the
horseshoe one reveals easier to translate into a circuit
representation of the \textsc{Cnot} gate.
Let us consider Fig. \ref{fig:horseshoe}. Since we realize our
computation within the one-way model, we perform simultaneous
single-qubit measurements on qubits $3$ and $4$ and on qubits $6$
and $1$ and then read the corresponding output on qubits $5$ and
$2$, both encoded in the polarization DOF.

We pointed out four possible measurement patterns in order to
accomplish different logical operations, depending on the bases
chosen for the single-qubit measurements. From now on, when
referring to a given measurement basis we will always think of the
so-called ``laboratory basis'' (LB), which differs from the
``cluster basis'' (CB) because of the presence of the local operations
affecting qubits $3$ ($X_3 \, \HH_3$) and $4$ ($\HH_4$) (see Eq.
(\ref{eq:clusterequiv})). The four considered measurement patterns, 
both in the cluster and in the laboratory bases, are listed in
Table \ref{CNOTcases}.

For each pattern a corresponding computational circuit can be derived.
In Fig. \ref{fig:circuits_caseI} we show the detailed derivation of the corresponding circuit
for the first considered pattern: the measurements implement the ``Cluster algorithm'' 
(see figure) and the change between the CB and the LB 
corresponds to the final gates (labeled as ``Change of basis'' in the figure). 
The circuit can be equivalently written as shown in the right part:
it consists of two single qubit rotations and a \textsc{Cnot} gate. The Pauli errors, as usual, depend on the
measurement results of qubits 3, 4, 6 and 1.
In Fig. \ref{fig:circuits} we show the equivalent circuits corresponding to the other three measurement patterns
we have considered.

By taking into account their circuit representations shown in 
Fig. \ref{fig:circuits_caseI} and Fig. \ref{fig:circuits}, we can write the expected output state, encoded in the
physical qubits 5 (photon $B$, control) and 2 (photon $A$, target), for each measurement pattern:
\begin{subequations}
\label{output}
\begin{align}
\ket{\psi_{out}}_{I} =& (X_{5}^{s_3 +s_6}Z_{2}^{s_1+ s_4} X_{2}^{s_3 +s_6})
\times\label{output:one} \\
&\times Z_5\textsc{Cnot}_{52} [R^{(5)}_x(\alpha) \otimes R^{(2)}_z(\beta)]{\ket{0}}_{5}{\ket{+}}_{2}, 
\notag\\
\ket{\psi_{out}}_{II} =& (X_{5}^{s_3 +s_6}X_{2}^{s_3 +s_4+s_6} )
\times\label{output:two} \\
&\times Z_5  \textsc{Cnot}_{52} [R^{(5)}_x(\alpha)  \otimes \openone_2]{\ket{0}}_{5}{\ket{0}}_{2},
\notag\\
\ket{\psi_{out}}_{III} =& (Z_{5}^{s_3} Z_{2}^{s_1 +s_4})
\times\label{output:three} \\
&\times Z_5 \, \textsc{Cnot}_{52}  [\openone_5\otimes R^{(2)}_z(\beta)]{\ket{+}}_{5}{\ket{+}}_{2}, 
\notag\\
\ket{\psi_{out}}_{IV} =& (Z_{B}^{s_3}X_{A}^{s_4}) Z_5 \, \textsc{Cnot}_{52} {\ket{+}}_{5}{\ket{0}}_{2},\label{output:four}
\end{align}
\end{subequations}
where $R_x(\alpha) = e^{-i\alpha X/2}$ corresponds to a
counterclockwise rotation through an angle $\alpha$ about the $x$
axis of the Bloch sphere (an analogous definition holds for
$R_z(\beta)$). The presenve of the single-qubit
Pauli errors $X$ and $Z$ depends on the measurement output of
the corresponding qubit (represented by $s_i$ for $i = 1, 3, 4,
6$, see the comment following Eq. (\ref{alphabasis})).
It comes out from the expressions of $\ket{\psi_{out}}$ in \eqref{output}
that the computations can be interpreted as single qubit 
transformations followed by a \textsc{Cnot}
gate acting on different input states. 
Precisely, by rewriting the computation in the error-free case we obtain:
\begin{subequations}
\label{output_simple}
\begin{align}
\ket{\psi_{out}}_{I} =& Z_5\textsc{Cnot}_{52} [R^{(5)}_x(\alpha) \otimes R^{(2)}_z(\beta)]{\ket{\psi_{in}}}_I, 
\label{simple:one}\\
\ket{\psi_{out}}_{II} =& Z_5  \textsc{Cnot}_{52} [R^{(5)}_x(\alpha)  \otimes \openone_2]{\ket{\psi_{in}}}_{II},
\label{simple:two}\\
\ket{\psi_{out}}_{III} =& Z_5 \, \textsc{Cnot}_{52}  [\openone_5\otimes R^{(2)}_z(\beta)]{\ket{\psi_{in}}}_{III}, 
\label{simple:three} \\
\ket{\psi_{out}}_{IV} =& Z_5 \, \textsc{Cnot}_{52} {\ket{\psi_{in}}}_{IV},\label{simple:four}
\end{align}
\end{subequations}
where the input states are ${\ket{\psi_{in}}}_I={\ket{0}}_{5}{\ket{+}}_{2}$,
${\ket{\psi_{in}}}_{II}={\ket{0}}_{5}{\ket{0}}_{2}$,
${\ket{\psi_{in}}}_{III}={\ket{+}}_{5}{\ket{+}}_{2}$ and
${\ket{\psi_{in}}}_{IV}={\ket{+}}_{5}{\ket{0}}_{2}$.

Let's start from pattern $IV$: in this case, by looking at the
measurement basis given in Table \ref{table:patterns}, it is 
possible to reinterpret the four tomographic reconstructions 
of the cluster state $\LCtilde$ with respect to the polarization DOF 
as a one-way computation (here the \textsc{Cnot} operation).
Precisely, the measurement of qubits 3 and 4 in the computational basis
corresponds to selecting different modes of the cluster. The
output is then encoded in the polarization of the two photons and the 
four Bell states correspond to the four different outputs of the computation.
In fact, it is easy to show that $\ket{\psi_{out}}_{IV}={\ket{\phi^-}}_{52}$ in
the error-free case. The other three Bell states are obtained by applying
the different Pauli errors.
Hence the tomographic reconstructions of the polarization states
given in Sec \ref{sec:char} suffice to the
experimental proof of the correct functioning of the realized
logic gate within the specific framework of pattern $IV$.

\begin{figure}[h!]
\begin{center}
\includegraphics[width=9cm]{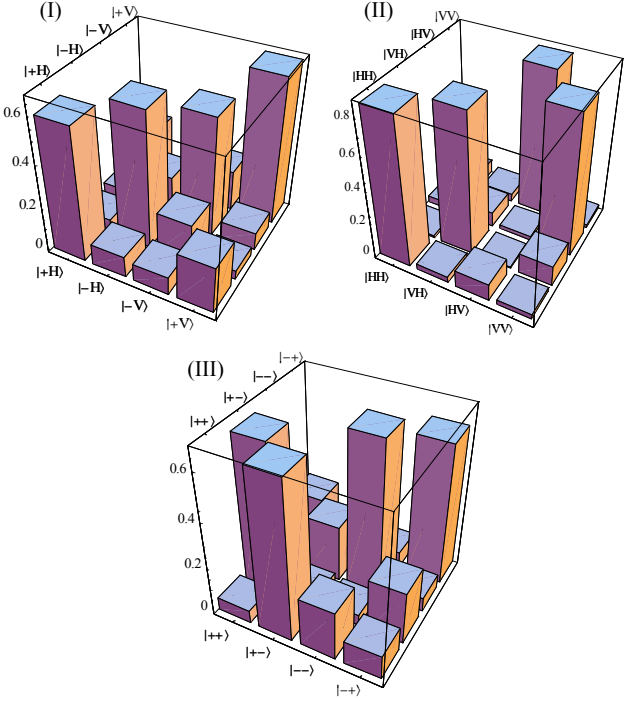}
\caption{Graphic representation of the
I-O matrices for the considered \textsc{Cnot} operation. The sublabels
indicate the pattern to which each matrix refers to. The sets of
input and output states are listed in Table \ref{IO:fidelities}
and can be read on the upper (input) and lower (output) axis.}
\label{CNOT:IO}
\end{center}
\end{figure}

Let's then consider patterns $I$, $II$ and
$III$ in the error-free case, which means that $s_i = 0$ for $i = 1, 3, 4, 6$.
Moreover we set $\alpha = \beta = 0$, implying that $R_x(0) = R_z(0) = \openone$.
These hypothesis lead to the output states $\ket{\psi_{out}}_{I}$,
$\ket{\psi_{out}}_{II}$ and $\ket{\psi_{out}}_{III}$, all in the
form of separable states of the two photons $A$ and $B$, and
establish a first set of input and output states for the three
cases. In these conditions, it is interesting to reconstruct the
input-output (I-O) matrices for the realized \textsc{Cnot} gate, whose
knowledge enables the further calculation of the fidelities
associated to the output states
(\ref{simple:one})-(\ref{simple:three}). The experimental results
are listed in Table \ref{IO:fidelities}, while Fig. \ref{CNOT:IO}
shows a graphic representation of the I-O matrices.
The fidelity values show that the gate built on the generated
cluster state $\LCtilde$ operates as expected.

\begin{table}
\caption{\label{IO:fidelities}Input-output states,
corresponding to the first three measurement patterns,
expressed in the polarization basis.
Here the ``AB'' ordering is used.
}
\begin{ruledtabular}
\begin{tabular}[b]{|cccc|}
Pattern & Input state \footnotemark[2] & Expected output state \footnotemark[2] & Fidelity \\
\hline
$I$ & $\ket{+H}_{AB}$ & $\ket{+H}_{AB}$ & $0.6052 \pm 0.0084$ \\
& $\ket{-H}_{AB}$ & $\ket{-H}_{AB}$ & $0.6657 \pm 0.0077$ \\
& $\ket{-V}_{AB}$ & $\ket{-V}_{AB}$ & $0.5476 \pm 0.0066$ \\
& $\ket{+V}_{AB}$ & $\ket{+V}_{AB}$ & $0.6223 \pm 0.0069$ \\
\hline
$II$ & $\ket{HH}_{AB}$ & $\ket{HH}_{AB}$ & $0.8716 \pm 0.0050$ \\
& $\ket{VH}_{AB}$ & $\ket{VH}_{AB}$ & $0.8348 \pm 0.0072$ \\
& $\ket{HV}_{AB}$ & $\ket{VV}_{AB}$ & $0.8710 \pm 0.0053$ \\
& $\ket{VV}_{AB}$ & $\ket{HV}_{AB}$ & $0.8376 \pm 0.0065$ \\
\hline
$III$ & $\ket{++}_{AB}$ & $\ket{+-}_{AB}$ & $0.6541 \pm 0.0111$ \\
& $\ket{+-}_{AB}$ & $\ket{++}_{AB}$ & $0.6798 \pm 0.0088$ \\
& $\ket{--}_{AB}$ & $\ket{--}_{AB}$ & $0.6741 \pm 0.0108$ \\
& $\ket{-+}_{AB}$ & $\ket{-+}_{AB}$ & $0.6096 \pm 0.0093$ \\
\end{tabular}
\end{ruledtabular}
\end{table}

\begin{figure}[!h]
\begin{center}
\includegraphics[width=9cm]{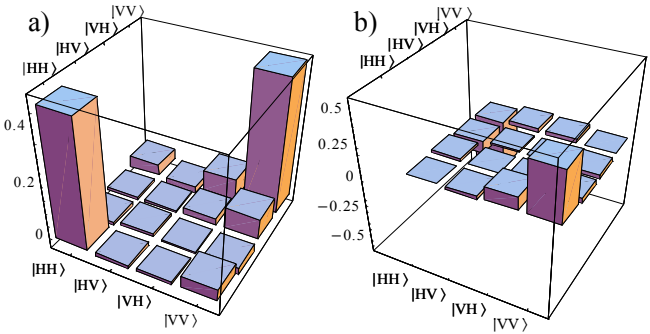}
\caption{Tomographic reconstruction of the
polarization entangled output state
$\ket{\psi_{out}^{\prime}}_{II}$. Both the (a) real and (b)
imaginary components are shown. The corresponding theoretical
state is $-\frac{1}{\sqrt{2}} (\ket{HH}_{AB} - i
\ket{VV}_{AB})$.} \label{CNOT:tomo}
\end{center}
\end{figure}

A further analysis consists of examining other possible values for
the rotation angles $\alpha$ and $\beta$ in the framework of an
error-free computation. By letting $\alpha$ and $\beta$ assume non-zero values 
during the computation, we
obtain other combinations of input and output product states. 
As an example, we consider the ``variant''
of pattern $II$ where $\alpha = 3 \pi/2$. We can then write the
output state as

\begin{equation}
\begin{split}
\ket{\psi_{out}^{\prime}}_{II} &= Z_5 \, \textsc{Cnot}_{52} R^{(5)}_x(3 \pi/2)
\ket{0}_{5}\ket{0}_{2} = \\ &= -\frac{1}{\sqrt{2}} (\ket{HH}_{AB}
- i \ket{VV}_{AB}).
\end{split}
\end{equation}
As we see, here we have an entangled two-photon state encoded in
polarization. When dealing with an entangled state of photons A
and B it is not possible to adopt an I-O matrix reconstruction
strategy in order to test the correctness of the gate's
functioning; it is now necessary to perform a tomographic
reconstruction of the output state corresponding to the considered
computation (this is exactly what happens with case $IV$, too).
The experimental tomographic analysis for the ``case $II$
variant'' is shown in Fig. \ref{CNOT:tomo}: the fidelity of the
output state $\ket{\psi_{out}^{\prime}}_{II}$ is $F = 0.879 \pm
0.017$.

\section{CONCLUSIONS AND PERSPECTIVES}
\label{sec:concl}
We have characterized the six-qubit linear cluster state $\LCtilde$, 
realized by starting from a two-photon state, hyperentangled in
 the three degrees of freedom of polarization and a double set 
of longitudinal momentum modes of the photons emitted over the 
degenerate cone of a Type I SPDC crystal. 

The importance of a six-qubit linear cluster state is twofold: while it represents a significant step in the 
research of quantum nonlocality, as recently demonstrated \cite{cecc09prl}, 
the realization of cluster states with an increasing number of qubits is important to the quantum computation community. 
Indeed, we have used this state by realizing the \textsc{Cnot} gate in the one-way quantum computation domain. 
For this purpose, the configuration chosen for the graph associated to the six-qubit state is that 
of a horseshoe (180$^\circ$ rotated) cluster state.

The \textsc{Cnot} results of our experiment are similar to those obtained within
the Hefei's group experiment \cite{gao09qph}, where a six-qubit graph state has been
created by using polarization and spatial modes of four photons. By that
technique new qubits encoded in different DOFs of the same photon are
 added by local operations. Multi-qubit entangled states realized by
this technique may find useful applications in one-way quantum computation.
 However, it has been already emphasized that in such states
polarization and longitudinal momentum of the same photon are not
independent \cite{vall09pra}. As a consequence, regarding their use in quantum
nonlocality tests, some problems may arise in the definition of
EPR's criterion for elements of reality \cite{gao09qph2}.

Both the existing approaches to one-way QC and error encoding, 
based on multi-photon and multi-DOF entanglement, contribute to make
 an all-optical architecture a serious contender for the ultimate 
goal of a large-scale quantum computer. However, scalable linear 
optics systems are required for the realization of more complex QC 
operations and algorithms. This is a very challenging objective, 
according to the current optical technology. One of the main reasons
 is that an increasing number of qubits requires the setup of bulk 
measurement systems of increasing complexity. At the same time, 
the need of an increasing number of qubits in a QC algorithm 
conflicts with the intrinsic limitations of the SPDC process.
Indeed, no more than few pairs of photons at a time are created
 by SPDC, due to its probabilistic nature. Moreover, multi-photon 
detection is seriously affected by the limited quantum efficiencies of modern detectors.

In order to take the maximum advantage of the possibilities offered
 by the current optical technology to increase power and speed of
 computational operations based on high dimension entangled photonic
 systems, we may conceive cluster states built on a number of
 photons entangled in many DOFs. Increasing the number of photons
 or encoding the qubits in other DOFs of the particles, besides
 polarization and longitudinal momentum, such as frequency, time 
bin and orbital angular momentum of the photons, are two complementary
 (but not exclusive) approaches to enhance the computational power and the information content.

It is worth to remember that increasing the number n of involved DOFs 
implies an exponential requirement of resources. For instance, $2^n$ $\bf k$-modes 
per photon must be selected within the emission cone to encode n qubits
 in each photon. However, according to 
the current optical technology, working with few DOFs (such as $n = 2,3,4$) offers 
still more advantages than working with a corresponding number of photon pairs, because 
of the higher repetition rate and state generation/detection efficiency. 
Indeed, by increasing the number of DOFs on which two photons are entangled, the overall 
detection efficiency and hence the repetition rate of detection is constant, since it scales 
as $\eta^N$, being $N$ the number of photons and $\eta$ the detector quantum efficiency, 
except for some factors depending on the measurement setup. Furthermore, an entangled state built 
on a larger number of particles is in principle more affected by decoherence because of the 
increased difficulty of making photons indistinguishable. In medium-term time 
scale a hybrid approach to QC (i.e. multi-DOF and multi-photon states) 
may represent a convenient solution to overcome the structural limitations in generation/detection of quantum photon states.

In view of an efficient linear optics quantum computation, the use
 of miniaturized optical circuits built on a chip in the realization
 of increasingly complex linear optical schemes consisting of many 
interferometers, whose feasibility has been recently demonstrated \cite{poli08sci,poli09sci,matt09npho,mars09ope}, 
is becoming of fundamental relevance. Indeed these new integrated 
structures guarantee high fidelities and highly intrinsic phase 
stability of the measurements necessary to perform the logical operations. 
Furthermore, the adoption of integrated optics may also enable the 
realization of novel kinds of multi-photon states. Hence, new exciting
 perspectives implying the solution of new problems are opened in the 
application of miniaturized optical structures with multi-photon multi-DOF entangled states.


\end{document}